\documentclass[conference]{IEEEtran}
\IEEEoverridecommandlockouts

\usepackage{amsfonts}
\usepackage{cite,graphicx,amsmath,amsthm}
\usepackage{amsmath}
\usepackage{xcolor}
\usepackage{subfig}
\usepackage{fancyhdr}
\usepackage{dsfont}
\usepackage{array,color}
\usepackage{bm}
\usepackage{float}
\usepackage{algorithm}
\usepackage{algpseudocode}
\usepackage{multirow}
\usepackage{booktabs}
\usepackage{multirow}
\usepackage{makecell}
\usepackage{tikz}
\usetikzlibrary{positioning} 
\usepackage{xcolor} 
\usepackage{hyperref}
\hypersetup{
    colorlinks=true,      
    linkcolor=black,      
    citecolor=black,      
    urlcolor=black,       
    anchorcolor=black,    
    filecolor=black,      
    menucolor=black,      
}
\DeclareUnicodeCharacter{200B}{\hspace{0pt}}
\usepackage{graphicx} 
\newcommand{\triangleq}{\stackrel{\triangle}{=}}
\usepackage{colortbl}
\makeatletter 
\newcommand*{\bibitem@shuaiWang}{blue}
\newcommand*{\bibitem@Geoffrion}{blue}
\makeatother 

\begin{document}
	
\title{ \huge Radio Map-Enabled 3D Trajectory and Communication Optimization for Low-Altitude Air-Ground Cooperation} 

\author{

\IEEEauthorblockN{
	Menghao Hu\IEEEauthorrefmark{1}, 
	Tong Zhang\IEEEauthorrefmark{3}, 
	Shuai Wang\IEEEauthorrefmark{4}, 
	Chiya Zhang\IEEEauthorrefmark{3}, 
	Changyang She\IEEEauthorrefmark{3}, 
	Gaojie Chen\IEEEauthorrefmark{8}, 
	Miaowen Wen\IEEEauthorrefmark{2}
        \thanks{This work was supported by the National Natural Science Foundation of China under
Grant 62371444, in part by the Open Research Fund of National Mobile Communications Research
Laboratory, Southeast University under Grant 2025D07. \newline \indent Our code is at \href{https://github.com/Joshua05160/Radiomap}{\textcolor{cyan}{https://github.com/Joshua05160/Radiomap}}.
}
}

\IEEEauthorblockA{
	\IEEEauthorrefmark{1}School of Information Science and Engineering, Southeast University, China\\
}

\IEEEauthorblockA{
	\IEEEauthorrefmark{3}Harbin Institute of Technology (Shenzhen), Shenzhen, China\\
}

\IEEEauthorblockA{
	\IEEEauthorrefmark{4}Shenzhen Institute of Advanced Technology, Chinese Academy of Sciences, Shenzhen 518055, China\\
}

\IEEEauthorblockA{
	\IEEEauthorrefmark{8}5GIC \& 6GIC, Institute for Communication Systems, University of Surrey, Guildford GU2 7XH, UK\\
}

\IEEEauthorblockA{
	\IEEEauthorrefmark{2}School of Electronic and Information Engineering, South China University of Technology, China\\
}
}

		\maketitle
		
		\begin{abstract}
			
Low-altitude economy includes the application of unmanned aerial vehicles (UAVs) serving ground robots. In this paper, we investigate the 3-dimensional (3D) trajectory and communication optimization for low-altitude air-ground cooperation systems, where mobile unmanned ground vehicles (UGVs) upload data to UAVs. Specifically, we propose a joint optimization algorithm to maximize the minimum sum-rate of UGVs while ensuring quality of service and navigation constraints. The proposed algorithm integrates a successive convex approximation (SCA)-penalty method for UGV-UAV scheduling,  an SCA-based approach for UGV transmit power control, and   a novel warm-start particle swarm optimization with cross mutation (WS-PSO-CM). The WS-PSO-CM exploits the convex optimization for a statistical channel model to initialize the particle swarm. Simulation results show that the proposed algorithm achieves a $45.8$\% higher minimum sum-rate compared to the state-of-the-art PSO-CM under the same iterations. This gain can be translated to reducing computational time by $46.7$\% of PSO-CM. Furthermore, our simulation results reveal that UAVs dynamically adjust trajectories to avoid interference by buildings, and maintain proximity to UGVs to mitigate path-loss. 
		
		\end{abstract}

		
		\section{Introduction}
 
The low-altitude economy refers to the emerging industry ecosystem centered around the use of low-altitude ($<$ 1000m) airspace for applications such as unmanned aerial vehicles (UAVs), air taxis, and general aviation  \cite{ref1, YangWCL, ref2}. A key application of low-altitude economy is the UAV service for ground robots, i.e., unmanned ground vehicles (UGVs) \cite{Chen,Wenjun, Menghao, Jianqiang, MorshedIoTJ}. For this application, UGVs are adept at acquiring data from ground targets, yet encountering obstacles in wireless communication due to the hindrance posed by buildings and terrain\cite{Chen}. For this reason, UAVs are leveraged to collect data from mobile UGVs at various locations from a higher altitude, thereby substantially enhancing the throughput of UGVs \cite{Chen,Wenjun, Menghao, Jianqiang,  MorshedIoTJ}. Therefore, it is necessary to study the joint optimization of 3-dimensional (3D) UAV trajectory and UGV communication in low-altitude air-ground cooperation systems. 
		
Existing studies on low-altitude air-ground cooperation systems were explored extensively.  For example, in \cite{Wenjun}, the authors explored trajectory optimization for UAV-UGV cooperative emergency networks, where both UAVs and UGVs facilitate communication with users and relay critical information from disaster zones to external locations. For UAV relaying problem, a rotary-wing UAV placement problem to serve the UGVs for sum-rate maximization was investigated in \cite{Menghao}, where the trajectories of UGVs are considered. In \cite{Jianqiang}, UAVs offered aerial perspectives and real-time data to improve the efficiency of UGV path planning. The authors of \cite{MorshedIoTJ} envisioned that low-altitude UAVs and a high-altitude platform (HAP) work together to support delay-sensitive, computationally intensive applications for ground devices. However, how to design 3D trajectories and communications of UAVs to efficiently collect data from UGVs, considering planned UGV paths, are still unexplored.

To fill this gap, a prerequisite is a 3D radio map for the interested area, in order to providing an environment-aware and reality-imitative way to stimulate the algorithms operating in reality \cite{10272348, ss, ZhangR, psoRadioMap}.
The authors of \cite{ss}  introduced a radio map-assisted UAV path planning algorithm designed to avoid jammers and enhance flight efficiency across various terrains. In \cite{ZhangR}, the authors  framed 3D path planning as a shortest path problem using radio maps, ensuring stable UAV communication and improved efficiency through grid quantification. The authors of \cite{psoRadioMap} investigated a UAV-assisted relaying network based on radio maps, employing particle swarm optimization (PSO) to maximize network throughput. On the other hand, a dynamic radio map accounts for time-variant transmitter locations, bringing additional difficulties in processing and optimization. In \cite{DongD}, the authors tackled the challenge of dynamic radio map-assisted multi-UAV target tracking by introducing grid-based and particle filter-based approaches. In \cite{PanTCOM}, PSO is combined with the genetic algorithm (GA) for 3D UAV trajectory optimization, accounting for obstacles. However, for dynamic radio map, specifically a large path-loss (PL) database instead of statistical channel models, the computational time of PSO with GA is prohibitively high,  which translates to unsatisfactory performance when computational time is constrained.



In this paper, for low-altitude air-ground cooperation systems, we therefore study joint optimization of 3D trajectory and communication in the presence of dynamic radio map.   
We decompose the mixed integer non-linear programming (MINIP) problem into subproblems, including SCA-penalty based scheduling, SCA-based power control, and warm start particles swarm optimization with cross mutation (WS-PSO-CM). Our WS-PSO-CM  enhances PSO-CM in \cite{PanTCOM} by the following novel idea: we initialize the particle swarm by the convex optimization results from statistical channel model. Thereby, follow-up PSO-CM is just evolving the convex optimization results for continuously enhancing the performance. Simulation results show $45.8\%$ in minimum sum-rate using the same  iterations, compared with PSO-CM in \cite{PanTCOM}. This gain translates to reducing computational time by $46.7$\%, if WS-PSO-CM is with the same performance of PSO-CM. As a useful insight, we find that UAVs tend to avoid interference by buildings and fly near UGVs to mitigate PL. 

		\section{System Model and Problem Formulation}
%
%

		We consider an air-ground cooperation system, where $N$ UGVs 
		communicate with $M$ UAVs.  We denote the 3D locations of UAVs and UGVs at time slot $t$ by 
			${\bm{\ell}}^{\text{\text{UAV}}}_{m}[t]=(\bm{\alpha}_{m}[t],H_{m}[t])=(x_{m}^{\alpha}[t], y_{m}^{\alpha}[t], H_{m}[t])\in \mathbb{R}^{1\times3}$ and ${\bm{\ell}}^{\text{UGV}}_{n}[t]=(\bm{\beta}_{n}[t],0)=(x_{n}^{\beta}[t], y_{n}^{\beta}[t], 0)\in \mathbb{R}^{1\times3}$, respectively, where $m \in \mathcal{M}=\{1,2,...,M\}$, $n \in \mathcal{N}=\{1,2,...,N\}$ and $\mathbb{R}^{1\times3}$ denotes a $1\times3$ vector in real number space. To avoid collision, the distance between UAVs should be larger than a minimum distance,
		\begin{equation}
			\label{anti-collision}\lVert{\bm{\ell}}^{\text{UAV}}_{m}[t]-{\bm{\ell}}^{\text{UAV}}_{m'}[t] \rVert_2 \geq d_{\min}, \,\,\, m \neq m',  \,\forall m,m',t.
		\end{equation}		
		 Also, we define a binary scheduling variable $a_{m,n}[t] \in \{0,1\}$, where if UGV $n$ communicates with UAV $m$ at time slot $t$  $a_{m,n}[t]=1$, and otherwise $a_{m,n}[t]=0$. Furthermore, at each time slot, one UAV only communicates with one UGV, i.e.,
		\begin{equation}
		\label{TDMA1}\sum_{n=1}^{N}a_{m,n}\left[t\right]\leq 1, ~ \sum_{m=1}^{M}a_{m,n}\left[t\right]\leq 1,\,\,\, \forall m,n,t. 
		\end{equation}

 The PL map for UGV $n$ as transmitter at time slot $t$ is defined as $\textbf{H}_{n}[t] \in \mathbb{R}^{X \times Y \times Z}$, where we partition the 3D physical space into $X \times Y \times Z$ cubes and each
			cube has a volume of $\delta^3$.  The PL in map $ \textbf{H}_{n}[t]$ at the 3D coordinate index $ (x_{m},y_{m},z_{m})$ is denoted by $h_{n,t}(\omega_m)$ with index $\omega_m = (x_{m},y_{m},z_{m})$. This index can be mapped to the actual position of UAV by $ x_{m}=\left\lfloor  (x_{m}^{\alpha}[t]-X_{\min})/\delta \right\rfloor +1$, $ y_{m} =\left\lfloor  (y_{m}^{\alpha}[t]-Y_{\min})/\delta \right\rfloor
			+1$, and $z_{m} =\left\lfloor (H_{m}[t]-H_{\min})/\delta \right\rfloor+1$, where $X_{\min}$, $Y_{\min}$ and $H_{\min}$ act as anchor points and $H_{\min}$ also denotes UAVs' minimum flight altitude. 
		 The rate between UGV $n$ and UAV $m$ at time slot $t$ is calculated as $
				 R_{m,n}\left[t\right] = \log_2\left(1+a_{m,n}[t] \text{SINR}_{m,n}[t]\right),
		$
		where { $\text{SINR}$ $=  \frac{h_{n,t}(\omega_m)P_{n}[t]}{\sum_{p\neq n}^{N}(\sum_{q=1}^{M}a_{pq}[t])h_{p,t}(\omega_m)P_{p}[t]+N_0}$,} with $N_{0}$ denoting the additive white Gaussian noise (AWGN) power.
		
		To maximize the minimal average sum-rate of UGVs under scheduling, power, quality of service and UAV flight constraints, we optimize the UGV-UAV scheduling set $\mathcal{A}=\{a_{m,n}\left[t\right]\}$, UAV 3D trajectory set $\mathcal{Q}=\{\bm{\alpha}_{m}[t]\}$ and UGV transmit power set $\mathcal{P}=\{P_{n}\left[t\right]\}$. Mathematically, this problem can be formulated as the following mixed-integer programming problem, \begin{subequations}		
			\begin{eqnarray}
		\!\!\!\!\!	(\text{P0})	 \,\,\,				 \max_{\mathcal{A},\mathcal{Q},\mathcal{P}}~&&  \!\!\!\!\!\!\!\!\! \min\frac{1}{T}\sum_{t=1}^{T}\sum_{m=1}^{M} R_{m,n}[t] \nonumber \\
				\textrm{s.t.}~~ 
				\nonumber&&\!\!\!\!\!\!\!\!\!{\eqref{anti-collision}}, {\eqref{TDMA1}},\\
				\label{qos}&&\!\!\!\!\!\!\!\!\! R_{m,n}[t]\geq a_{m,n}[t]R_{\min},  \forall m,n,t,\\
				\label{max-speed}&&\!\!\!\!\!\!\!\!\! \lVert{\bm{\ell}}^{\text{UAV}}_{m}[t+1]-{\bm{\ell}}^{\text{UAV}}_{m}[t] \rVert_2 \leq V_{\max}\tau,t <T, \\
			  	\label{HeightC}&&\!\!\!\!\!\!\!\!\! H_{\min} \leq H_{m}[t] \leq H_{\max}, \forall m, t,\\
			  	\label{angle}&&\!\!\!\!\!\!\!\!\!  {\theta_{m}[t] \leq \theta_{\max}, \forall m, t,}\\
			  	\label{building}&&\!\!\!\!\!\!\!\!\! { \bm{\ell}^{\text{UAV}}_{m}[t] \notin \mathcal{B}, \forall m, t,}\\
				\label{power}&&\!\!\!\!\!\!\!\!\! 0 \leq P_{n}[t] \leq P_{\max}, \forall n,t,\\
				\label{0-1}	&&\!\!\!\!\!\!\!\!\! a_{m,n}[t] \in\left\{0,1\right\}, \forall m,n,t,	 
			\end{eqnarray}			\label{P0}
		\end{subequations}
        $\!\!\!$where {\eqref{qos}} is the quality of service constraint for each communicating link, with $R_{\min}$ denoting the minimum rate;  {\eqref{max-speed}} ensures the speed of UAVs is not more than $V_{\max}$ and $\tau$ is the duration of each time slot; {{\eqref{HeightC}} ensures the flying altitude of UAVs is between $H_{\max}$ and $H_{\min}$}; {\eqref{angle} ensures UAVs' turning angle does not exceed the maximum turning angle limit $\theta_{\max}$, where $\theta_{m}[t] = \angle({\bm{\ell}}^{\text{UAV}}_{m}[t]-{\bm{\ell}}^{\text{UAV}}_{m}[t-1],{\bm{\ell}}^{\text{UAV}}_{m}[t+1]-{\bm{\ell}}^{\text{UAV}}_{m}[t])$. {\eqref{building}} ensures that UAVs do not collide with buildings, where $\mathcal{B}$ denotes the 3D building  space.} {\eqref{power}} ensures UGVs transmit power is not more than $P_{\max}$.

\begin{figure}[t]
			\centering
			\includegraphics[width=0.65\linewidth]{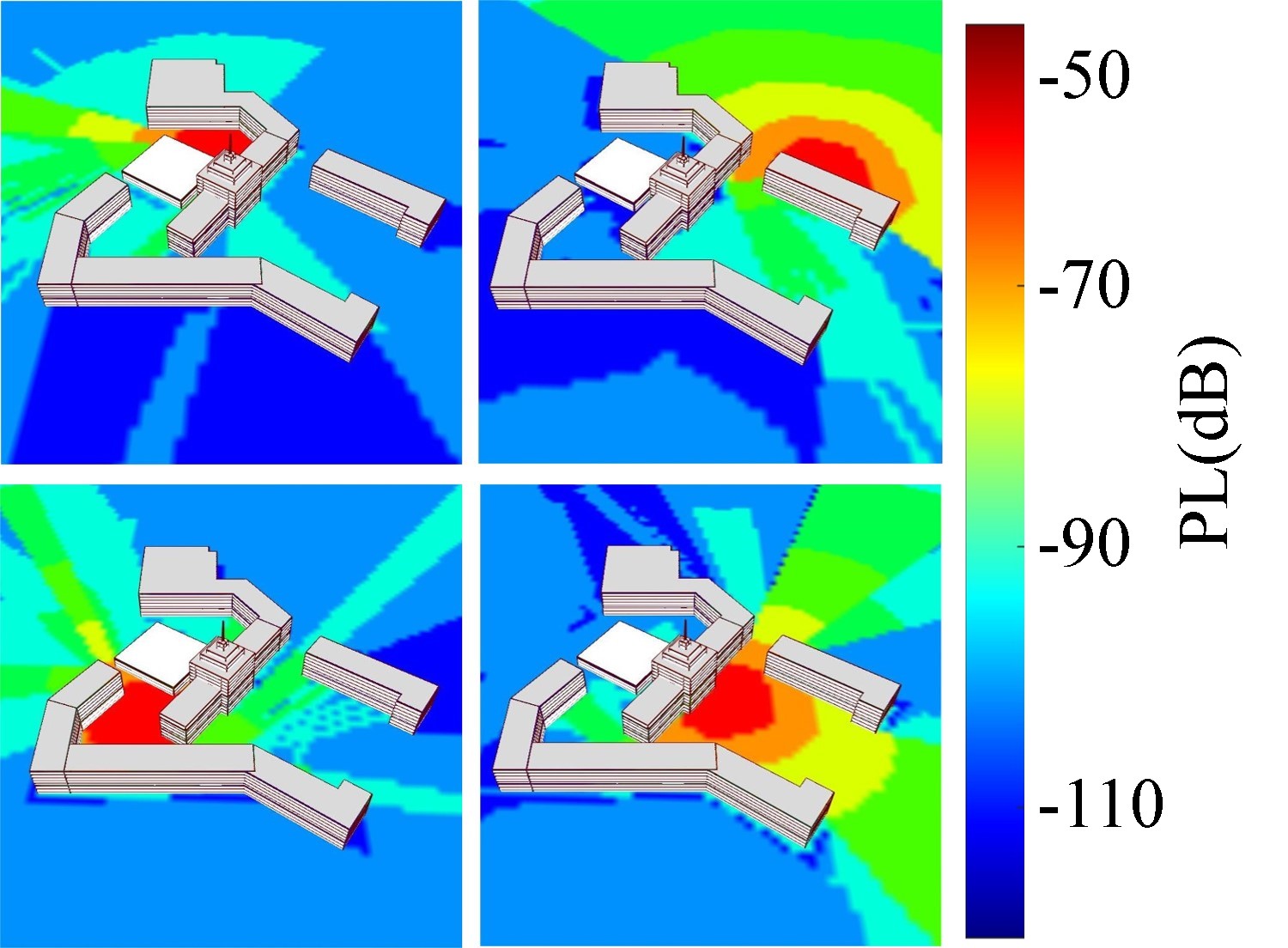}
			\captionsetup{justification=raggedright, singlelinecheck=off} 
			\caption{HITSZ PL maps at different UGV locations, made by \textsc{Ranplan Academic V6.8.0} {\small{https://www.ranplanwireless.com}}}
			\label{HIT}
		\end{figure}

		We can establish a digital twin model of {Harbin Institute of Technology, Shenzhen (HITSZ)} by \textsc{Ranplan Academic V6.8.0} and obtain the PL map database, where Fig.~\ref*{HIT}  shows the PL heat map for UGVs at different positions. Next, we design an algorithm on general radio
		maps. This algorithm will be validated via the above radio map database in simulation.
 
		\section{Proposed Algorithm}
Unfortunately, solving Problem (P0) in its entirety is challenging, due to MINLP problem. {Therefore, we decompose Problem $(\text{P0})$ into  UGV-UAV scheduling optimization subproblem,  UGV transmit power optimization subproblem, and UAV 3D trajectories optimization subproblem \cite{TongZ}.} 
		\subsection{Scheduling Optimization}
		Given the designed trajectory and power control, we consider to optimize the UGV-UAV scheduling problem. {We can transform the binary constraint {\eqref{0-1}} to a linear constraint  {\eqref{continuous}} with a non-convex penalty term $\psi(\mathcal{A})=\eta\sum_{t=0}^{T}\sum_{n=1}^{N}\sum_{m=1}^{M}a^{2}_{m,n}[t]-a_{m,n}[t]$ to ensure the variables tending $0$ or $1$. $\eta$ is selected to align with the order of magnitude of the objective function \cite{Geoffrion}, }
		\begin{equation}
			\begin{aligned}
				\label{continuous}	0\leq a_{m,n}[t]\leq 1, \quad\forall m,n,t.
			\end{aligned}
		\end{equation}
	Then, we have 
		\begin{eqnarray}
			(\text{P1}) \,\,\,\, \nonumber\max_{\mathcal{A},~\mu} && \mu+\psi(\mathcal{A}) \nonumber \\
			\text{s.t.} && {\eqref{TDMA1}}, {\eqref{qos}}, {\eqref{continuous}}, \nonumber \\	
			&& \frac{1}{T}\sum_{t=1}^{T}\sum_{m=1}^{M} R_{m,n}[t]\geq \mu, \, \forall n, \label{lamda}
		\end{eqnarray}
	{where $\max\,\, \min$ problem is equivalent to $\max$ problem by introducing constraint \eqref{lamda} and variable $\mu$.} $R_{m,n}[t]$ is then rewritten as  $ \log_{2}  (a_{m,n}[t]P_{n}[t]h_{n,t}(\omega_m)+\mathcal{I}_{m,n}[t]+N_{0} ) - \log_{2} (\mathcal{I}_{m,n}[t]+N_{0} )$, where the first and second terms are denoted by $\widehat{R}_{m,n}[t]$ and $\widetilde{R}_{m,n}[t]$, respectively, and $\mathcal{I}_{m,n}[t] =\sum_{p\neq n}^{N}(\sum_{q=1}^{M}a_{pq}[t])h_{p,t}(\omega_m)P_{p}[t]$.	
\begin{figure*}[t]
	\begin{equation}
		\begin{aligned}
			\label{B}&\widetilde{R}_{m,n}[t]\leq B^{ub}_{m,n}[t] \triangleq \log_2(\sum_{p \neq n}^{N} (\sum_{q \neq m}^{M} a^{l}_{q,p}[t]) h_{p,t}(\omega_m) P_{p}[t] + N_0 )+ \frac{h_{p,t}(\omega_m) P_{p}[t]\left(\sum_{q \neq m}^{M} (a_{q,p}[t] - a^{l}_{q,p}[t])\right)\ln2}{\sum_{p \neq n}^{N} \left(\sum_{q \neq m}^{M} a^{l}_{q,p}[t]\right) h_{p,t}(\omega_m)P_{p}[t] + N_0}, 
		\end{aligned}
	\end{equation}
	\begin{equation}
		\begin{aligned}
			\label{C}	&\widetilde{R}_{m,n}[t]\leq C^{ub}_{m,n}[t] \triangleq \log_2(\sum_{p\neq n}^{N}(\sum_{q\neq m}^{M}a_{q,p}[t])h_{p,t}(\omega_m)P^{l}_{p}[t]+N_0)+\frac{(\sum_{q\neq m}^{M}a_{q,p}[t])h_{p,t}(\omega_m)(P_{p}[t]-P^{l}_{p}[t])\ln2}{\sum_{p\neq n}^{N}(\sum_{q\neq m}^{M}a_{q,p}[t])h_{p,t}(\omega_m)P_{p}[t]+N_0},
		\end{aligned}
	\end{equation}
	\hrule
\end{figure*}
		Unfortunately, $(\text{P1})$ is non-convex, because ${\eqref{qos}}$ and ${\eqref{lamda}}$ are the difference of two concave functions w.r.t  $a_{m,n}[t]$.
		Thus, we leverage SCA technique to perform first-order Taylor expansion on $ \widetilde{R}_{m,n}[t]$ at $a^{l}_{q,p}[t]$ to obtain its linear upper bound in ${\eqref{B}}$, shown on the top of the next page, where $l$ denotes the $l$th iteration. 
		Now the non-convexity of the problem stems from the added penalty term. Thus, we conduct first-order Taylor expansion  $ \psi(\mathcal{A})\geq \psi^{lb}(\mathcal{A})=\eta\sum_{t=0}^{T}\sum_{n=1}^{N}\sum_{m=1}^{M}\left(-(a^{l}_{m,n}[t])^{2}+(2a^{l}_{m,n}[t]+1)a_{m,n}[t]\right)$ for the penalty term. Finally, the SCA-form of Problem $(\text{P1})$ in the $l$th iteration is formulated as follows,
		\begin{subequations}
			\label{P_4}
			\begin{eqnarray}
			\!\!\!\!\!\!\!\!	(\text{P1})_l\,\,  
				\max_{\mathcal{A},~\mu}~&& \!\!\!\!\!\!\!\!\!\!\!\! \nonumber \,\, \mu+\sum_{t=1}^{T}\sum_{n=1}^{N}\sum_{m=1}^{M} \psi^{lb}(\mathcal{A})\\
			\!\!\!\!\!\!\!\!	\textrm{s.t.}~~ 
				\nonumber&&\!\!\!\!\!\!\!\!\!\!\!\! {\eqref{TDMA1}}, {\eqref{continuous}},\\
			\!\!\!\!\!\!\!\!	&&\!\!\!\!\!\!\!\!\!\!\!\! \frac{1}{T}\sum_{t=1}^{T}\sum_{m=1}^{M} \left(\widehat{R}_{m,n}[t]-B^{ub}_{m,n}[t]\right)\geq \mu,  \forall n,\\	
			\!\!\!\!\!\!\!\!	&&\!\!\!\!\!\!\!\!\!\!\!\! \widehat{R}_{m,n}[t]-B^{ub}_{m,n}[t]\geq a_{m,n}[t]R_{\min},  \forall m,n,t.
			\end{eqnarray}
		\end{subequations}		
Problem $(\text{P1})_l$ is convex, which can be solved by CVX.  
{By iteratively solving Problem $(\text{P1})_l$, a solution to Problem $(\text{P1})$ is obtained, which could be KKT points for some problems. }
		\subsection{Transmit Power Optimization}
		Given the designed trajectory and a scheduling policy, UGV transmit power optimization subproblem $(\text{P2})$ is formulated as 
		\begin{equation*}
			(\text{P2})\,\,\,\max_{\mathcal{P},~\mu}\,  \mu \,\,\,\,\,\, 
		{\text{s.t.}   \quad{\eqref{qos}},   {\eqref{power}},  \eqref{lamda}.		}	 
		\end{equation*}

		Then we perform first-order Taylor expansion on $\widetilde{R}_{m,n}[t]$ in ${\eqref{C}}$, shown on the top of this page.		
		After that, we have the following SCA-form of Problem $(\text{P2})$ in the $l$th iteration,
		\begin{subequations}
			\label{P_2'}
			\begin{align}
			\!\!\!\! 	(\text{P2})_l\,\,
				\max_{\mathcal{P},~\mu}~&\nonumber \,\,\mu\\
			\!\!\!\! 	\textrm{s.t.}~~ 
				\nonumber& {\eqref{power}},\\
			\!\!\!\! 	&\frac{1}{T}\sum_{t=1}^{T}\sum_{m=1}^{M} \left(\widehat{R}_{m,n}[t]-C^{ub}_{m,n}[t]\right)\geq \mu,  \forall n,\\	
		\!\!\!\!\ 		&\widehat{R}_{m,n}[t]-C^{ub}_{m,n}[t]\geq a_{m,n}[t]R_{\min},  \forall m,n,t,
			\end{align}
		\end{subequations}		
which is convex and can be solved by CVX. 
{By iteratively solving Problem \((\text{P2})_l\), a solution to Problem $(\text{P2})$ is obtained, which could be KKT points for some problems.}
		\subsection{{3D} Trajectories Optimization}
	Discrete PL map makes the gradient of objective function at UAV positions either zero or infinitely large. Thus, continuous optimization methods are incapable of solving this problem. To address this, we consider to adopt PSO \cite{PSO}. However, due to the curse of dimensionality and inherent nature of PSO, it is prone to getting trapped in unsatisfactory points\cite{PSO}. Nevertheless, we design a WS-PSO-CM, which 1) initiates at a good point and 2) facilitates diversity by cross and mutation.
	
	\textit{\underline{Step 1}: Warm-Start Particles Initialization}.
	We consider replacing the radio map with line-of-sight (LoS) channel model for a good initial point. According to {\cite{Model}}, LoS model reads $h_{m,n}[t]=\frac{P_{n}[t]L_{0}}{\lVert{\bm{\ell}}^{\text{UAV}}_{m}[t]-{\bm{\ell}}^{\text{UGV}}_{n}[t] \rVert_2^2}$,
	where $ L_0$ is the average path loss at reference distance $1$ m.  Subsequently, we obtain a continuous optimization problem $(\text{P3})$, aiming for optimizing 3D UAV trajectories.
	\begin{eqnarray}
		(\text{P3}) \,\,\, \max_{\mathcal{Q},~\mu} \mu\quad  \,\,\,
		\text{s.t.} \,\,\,  {\eqref{anti-collision}}, {\eqref{qos}}, {\eqref{max-speed}}, {\eqref{HeightC}}, {\eqref{angle}}, {\eqref{building}}, \eqref{lamda}. \nonumber  
	\end{eqnarray}

In regard to Problem $(\text{P3})$, we decompose the 3D trajectory optimization into horizontal and vertical trajectory optimization subproblems separately. Then, we adopt SCA for each subproblems, and iteratively solve them until convergence. This yields the so-called \textit{initial trajectory particles}. 

	\textit{\underline{Step 2}: Intelligent Searching with PSO-CM}.
	Hereafter, each particle denotes a trajectory in the solution space. The $ k$th particle is denoted by $ \mathcal{Q}_{k}$ and its velocity by $\mathcal{V}_{k}$, where 
$\mathcal{Q}_{k}=\left\{(\bm{\alpha}_{1}[1],H_{1}[1]), \cdot\cdot\cdot, (\bm{\alpha}_{M}[T],H_{M}[T]) \right\}, 
	 		\mathcal{V}_{k}=\left\{(v_{x}, v_{y},v_{z})_{[1,1]}, \cdot\cdot\cdot, (v_{x}, v_{y},v_{z})_{[M,T]} \right\}.
$
Particles update their velocity using their current velocity, global best, and local best and then update their position accordingly. The particle positions and velocity update strategy are given as follows.
	\begin{subequations}
		\begin{align}
			&\mathcal{V}_{k}=\omega \times\mathcal{V}_{k} + h_{1}\times\textsc{rand}_{1}\times\left(pBest_{k}-\mathcal{V}_{k}\right)\label{upadteV}\\
			&\phantom{\mathcal{V}_{k}=}+h_{2}\times\textsc{rand}_{2}\times\left(gBest-\mathcal{V}_{k}\right),\notag\\
			&\mathcal{Q}_{k}=\mathcal{Q}_{k}+\mathcal{V}_{k}\label{upadteQ},
		\end{align}
	\end{subequations}	
	where $ \omega$, $ h_{1}$ and $ h_{2}$ are inertial weight, cognitive parameter and social parameter, respectively, affecting the updating of position and velocity of the particles; $pBest_k$ and $ gBest$ are the $k${th} particle's position and the best particle's position at current iteration; $ \textsc{rand}_{1}$ and $ \textsc{rand}_{2}$ are two random variables ranging from $[0,1]$ with uniform distribution.
	
	During the UAV flight, besides system sum-rate, it is imperative to also consider 1) the UAV's flight speed constraint; 2) the maximum turning angle constraint {\eqref{angle}}, and 3) the collision constraints with buildings {\eqref{building}}. We therefore formulate the following fitness function, $F(\mathcal{Q}_{k}) = \alpha \Omega(\mathcal{Q}_{k})+\beta S(\mathcal{Q}_{k})+ \gamma A(\mathcal{Q}_{k})+\kappa C(\mathcal{Q}_{k})$,
	where $ \alpha$, $\beta$, $\gamma$, and $\kappa$ are the weight coefficients; 
{$\Omega(\mathcal{Q}_{k})=\min\sum_{t=1}^{T}\sum_{m=1}^{M} R_{m,n}[t]$ denotes the min sum-rate}; $S(\mathcal{Q}_{k})$ denotes the sum of the normalized velocities exceeding the maximum speed limit $V_{\max}$, i.e., $S(\mathcal{Q}_{k}) = \sum_{m=1}^{M}\sum_{t=2}^{T} \max (0, \frac{v_{m}[t] - V_{{\max}}}{V_{{\max}}} )$
	with {$ v_{m}[t] = \frac{\lvert {\bm{\ell}}^{\text{UAV}}_{m}[t]-{\bm{\ell}}^{\text{UAV}}_{m}[t-1]\rvert}{\tau}$};
  $ A(\mathcal{Q}_{k})$ denotes the sum of the normalized turning angle exceeding the maximum {limit}, i.e., $A(\mathcal{Q}_{k}) = \sum_{m=1}^{M}\sum_{t=2}^{T-1} \max (0, \frac{\theta_{m}[t] - \theta_{{\max}}}{\theta_{{\max}}} )$.  
  $ C(\mathcal{Q}_{k})$ denotes the sum of the amount of height of path points located within $\mathcal{C}_{b}$ exceeding the height of the building, i.e. $				C(\mathcal{Q}_{k}) = \sum_{m=1}^{M}\sum_{t=1}^{T-1}\max\left(0, \Delta_{m}[t]\right)$, 
		and $\Delta_{m}[t] =H_{m}[t]-H_{b}$ if ${\bm{\ell}}^{\text{UAV}}_{m}[t] \in \mathcal{C}_{b}$; otherwise $\Delta_{m}[t] =0$, 
	with $\mathcal{C}_{b}$ denoting the planar region of the buildings.	

	We incorporate crossover and mutation concepts from GA \cite{MIT} to enhance particle diversity in PSO.  
	\underline{Cross}: In each iteration, for particle $k$, set $\mathcal{Q}_{k} = \textsc{rand}_3\times \mathcal{Q}_{k} + (1 - \textsc{rand}_3)\times \mathcal{Q}_{k'}$ if $\textsc{rand}_4 \leq \sigma_c$; otherwise, $\mathcal{Q}_{k}$ remains unchanged, where $\mathcal{Q}_{k'}$ is another particle, $\sigma_c$ is the cross rate, and $\textsc{rand}_3$, $\textsc{rand}_4$ are random variables akin to $\textsc{rand}_1$.  
	\underline{Mutation}: For particle $k$, update $\mathcal{Q}_{k} = \mathcal{Q}_{k} + \mathbf{rand}_5 \times v_{\max}$ if $\textsc{rand}_6 \leq \sigma_m$; otherwise, $\mathcal{Q}_{k}$ stays the same, where $\mathbf{rand}_5$ is a matrix (entries uniformly in $[-1,1]$) matching $\mathcal{Q}_{k}$'s dimensions, $v_{\max}$ is the maximum velocity per dimension, and $\textsc{rand}_6$, $\sigma_m$ are the random variable and mutation rate in $[0,1]$, respectively.

	\begin{algorithm}[t]
	\label{algorithm1}\caption{Proposed Algorithm for Solving Problem $(\text{P0})$}
	\begin{algorithmic}[1]
		\State{}\textbf{Initialize}: $\mathcal{A}^{0}$, $\mathcal{P}^{0}$ and $\mathcal{Q}^{0}$; set $\text{iter}_1 = \text{iter}_2 = \text{iter}_3 = 1$, stopping accuracy $\epsilon > 0$; {and $\eta = 0.5$};
		\State\textbf{Repeat}:
		\State \quad\textbf{Repeat}:		
		\State \quad \quad Yield $\mathcal{A}^{\text{iter}_1}$ by solving $(\text{P1})_{\text{iter}_1}$ with $\mathcal{A}^{\text{iter}_1\text{-}1}$;
		\State \quad \quad Set $\text{iter}_1 \leftarrow \text{iter}_1+1$;	
		\State \quad\textbf{Until}: The fractional increase of $(\text{P1})_l$ is below $\epsilon$;
		\State \quad Solve Problem $(\text{P3})$ by warm start to initialize \{$\mathcal{Q}_k$\};
		\State \quad \textbf{Repeat}:
		\State \quad \quad Update particles' velocities according to {\eqref{upadteV}};
		\State \quad \quad Update particles' positions according to {\eqref{upadteQ}};
		\State \quad \quad Update the ${pBest_{k}}$;
		\State \quad \quad Select ${gBest}$ from ${pBest_{k}}$;
		\State \quad \quad Particles cross and mutate according to \underline{Cross} and \underline{Mutation} strategy;
		\State \quad \quad Set $\text{iter}_3\leftarrow \text{iter}_3+1$;
		\State \quad \textbf{Until}: $\text{iter}_3 = P_{\text{iter}}$;
		\State\textbf{Until}: The fractional increase of sum-rate 
		is below $\epsilon$;
		
		\State \textbf{Repeat}:		
		\State \quad Yield $\mathcal{P}^{\text{iter}_2}$ by solving $(\text{P2})_{\text{iter}_2}$ with $\mathcal{P}^{\text{iter}_2\text{-}1}$;
		\State \quad Set $\text{iter}_2 \leftarrow \text{iter}_2+1$;	
		\State \textbf{Until}: The fractional increase of $(\text{P2})_l$ is below $\epsilon$;
		\State \textbf{Output}: $\mathcal{A}^{\text{iter}_1}$, $\mathcal{P}^{\text{iter}_2}$ and $gBest$.
		\vspace{-0.15cm}
	\end{algorithmic}
\end{algorithm}
		\begin{figure*}[t]
		\centering
		\subfloat[]{\includegraphics[width=0.45\textwidth]{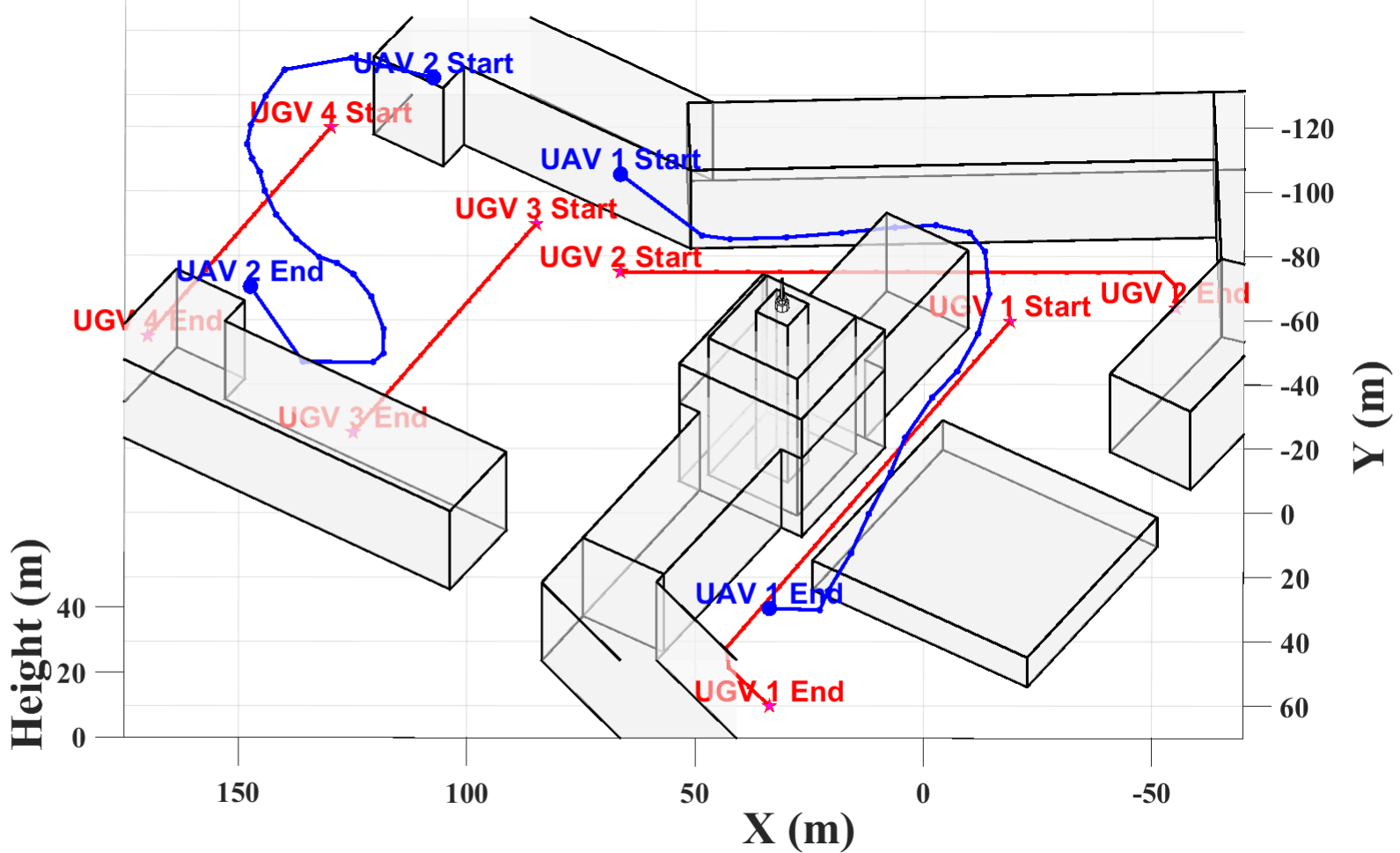}} \hfil
		\subfloat[]{\includegraphics[width=0.45\textwidth]{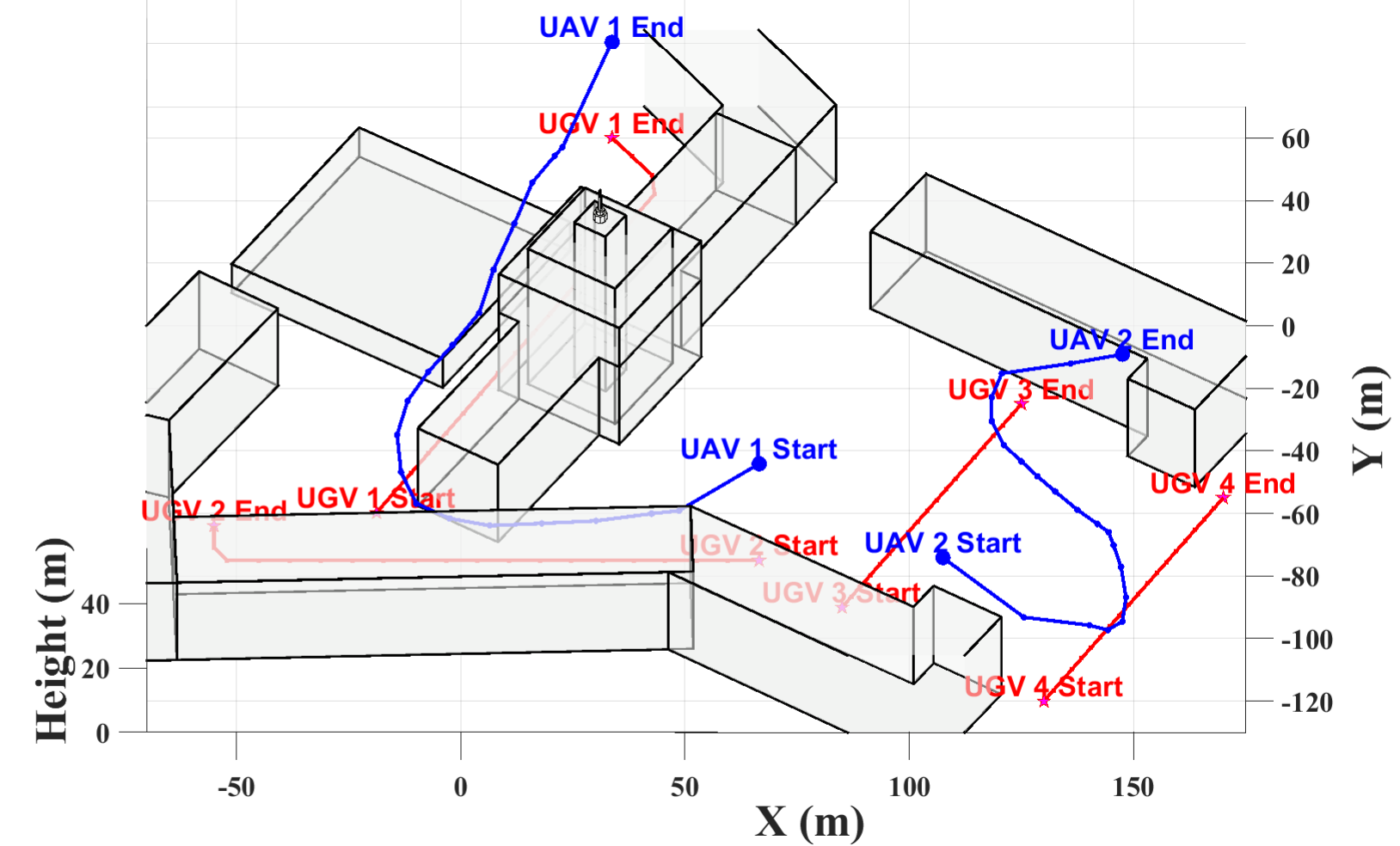}}
		\captionsetup{justification=raggedright, singlelinecheck=off}
		\caption{(a) High-altitude view from front; (b) High-altitude view from back.}
		\label{trajectory} 
	\end{figure*}		
		\subsection{Proposed Overall Algorithm and Complexity Analysis}
	Our proposed overall algorithm for solving Problem $(\text{P0})$ is given in Algorithm 1. 
{$\eta = 0.5$ is selected to ensure that the penalty term's order of magnitude matches that of the objective function. To prevent prematurely constraining the scheduling solution space, we postpone UGV transmission power optimization $\textsc{P2}$ until after completing the iterative scheduling $\textsc{P1}$ and trajectory planning $\textsc{P3}$ process.}
{The worst-case computational complexity of Algorithm 1 is $\mathcal{O}\bigl(K_1(MNT)^{3.5} + K_2 M^{1.5}(NT)^{3.5} + K_3 N^{1.5}(MT)^{3.5} + MT P_\text{iter} P_\text{num}\bigr)$.  
This comes from the following.  
Solving a convex problem with CVX interior point method has complexity $\mathcal{O}((V+C)^{1.5}V^2)$, where $V$ and $C$ are the number of variables and constraints, respectively.  
Solving Problem $(\text{P1})_l$ takes $\mathcal{O}(K_1(MNT)^{3.5})$.  
Solving Problem $(\text{P2})_l$ takes $\mathcal{O}(K_2 M^{1.5}(NT)^{3.5})$.  
The trajectory optimization (WS) takes $\mathcal{O}(K_3 N^{1.5}(MT)^{3.5})$.  
Here, $K_1$, $K_2$, and $K_3$ are number of iterations for accuracy $\epsilon$, given by $\frac{\log(c/t^0\gamma)}{\log(\epsilon)}$, where $c$ is the constraint count, $t^0$ is the initial point, $0<\gamma\leq 1$ is the stopping criterion, and $\epsilon$ is the stopping accuracy.}
 The computational complexity of PSO-CM is $\mathcal{O}(MTP_\text{iter}P_\text{num})$, where $P_\text{iter}$ and $P_\text{num}$ are the number of iteration and number of particles in the proposed WS-PSO-CM, respectively.

		\section{Simulation}

\subsection{Simulation Setup}		
	 Our simulation scenario is the teaching building area and hall of HITSZ, occupying the space of $240 \,\text{m} \times 400 \,\text{m} \times60 \,\text{m}$. Table \ref*{setup} presents the simulation parameter settings. 
	 The same digital twin is constructed in mathematical software for numerical calculation thereafter based on PL map data. The speeds of UGV $1 \,\&\, 2$ and UGV $3 \,\&\, 4$ are  $27.0$ km/h and $16.2$ km/h, respectively. The weight coefficients $ \alpha$, $\beta$, $\gamma$, and $\kappa$ are set $0.5$, $2$, $5$ and $5$, respectively. 
 
To validate the advantage of the proposed algorithm over the state-of-the-art, baseline algorithms are listed as follows:
		\begin{itemize}
			\item 	\textbf{\textsc{PSO-CM}:} No warm-start version of our WS-PSO-CM for Problem $(\textsc{P}3)$; The version is also that in \cite{PanTCOM}. 
			\item  \textbf{\textsc{PSO}:} No warm-start, cross and mutation version of our WS-PSO-CM for Problem $(\textsc{P}3)$; Note that this is also the standard version of PSO, and used in \cite{psoRadioMap}. 
			\item \textbf{\textsc{GA}:} Genetic algorithm\cite{MIT} for solving Problem $(\textsc{P}3)$; Note that the key of GA is the cross and mutation.
			\item 	\textbf{\textsc{RR}:} Scheduling UGVs in Round Robin manner for Problem $(\textsc{P}1)$; Note that Round Robin is the fairest scheduling scheme, without any focus on sum-rate.
			\item 	\textbf{\textsc{FP}:} Fix the UGV transmit power to $P_{\max}/2$ for Problem $(\textsc{P}2)$, while other variables are solved by Algorithm 1. This baseline is to test the energy consumption saving by optimization via our proposed algorithm.
		\end{itemize}
	
\subsection{Simulation Results}
	

	
\begin{figure}[t]
	\centering
	\includegraphics[width=2.35in]{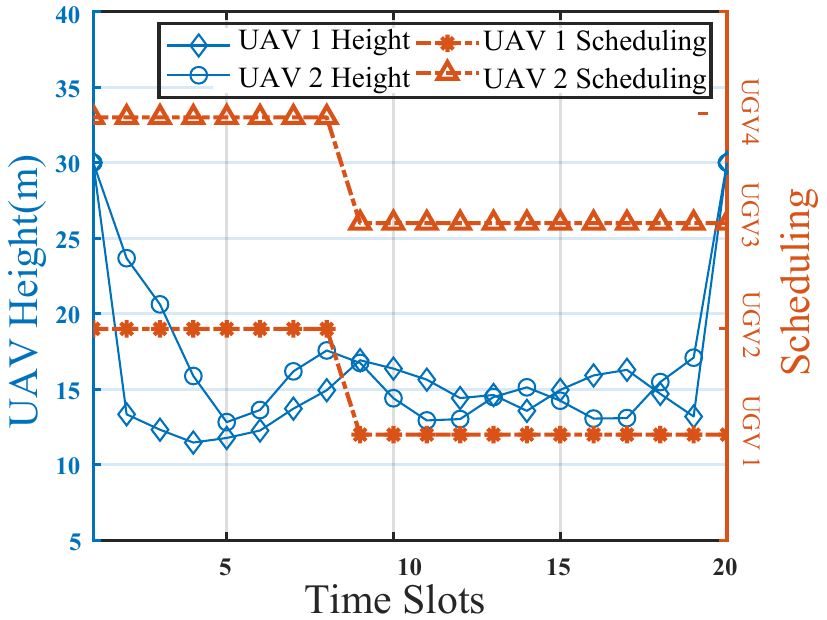}
	\caption{Optimized UAV height \& scheduling by our algorithm.} \label{F3a}
\end{figure}

Fig.~\ref*{trajectory} shows that the UAVs maintain proximity to UGVs while traversing through building shadow regions during their flights. The altitude variations observed in Fig.~\ref*{trajectory} and Fig.~\ref*{F3a} indicate that UAVs maintain a low flight altitude to minimize PL and reduce interference from other UGVs using building obstructions.  Furthermore, Fig.~\ref*{F3a} also shows that the UAVs ensure communication fairness among the UGVs during flight, since the optimized UGV scheduling are approximately balanced across 4 UGVs.
\begin{table}[t]
	\centering
	\caption{{Simulation setup}} \label{setup}
	\begin{tabular}{|l|l|l|}
		\hline
		\rowcolor{gray!20} \textbf{Variables} & \textbf{Interpretation} &\textbf{Setup} \\ \hline
		\( M \)   & Number of UAVs    & 2\\ \hline
		\( N \)   & Number of UGVs     & 4\\ \hline
		\( T \)   & Number of time slots       & 20\\ \hline
		\( P_\text{num} \) & Number of particles & 100\\ \hline
		\( N_0 \) & Power of AWGN  & -120 dBm \\ \hline
		\( L_0 \) & Reference pathloss  & 30 dB \\ \hline
		\( P_\text{iter} \) & Maximal number of iterations & 100\\ \hline
		\( \sigma_{c} \) & Cross rate & 0.1\\ \hline      
		\( \sigma_{m} \) & Mutation rate & 0.1\\ \hline 
		\( \delta \) & Cube side length    & 5 m \\ \hline
		\( \theta_{\max} \) & Max turning angle & \(40^\circ\) \\ \hline 
		{\( d_{\min} \)} & Anti-collision distance  & 10 m \\ \hline
		\( V_{\max} \) & UAV maximum speed & 20 m/s \\ \hline
		\( R_{\min} \) & Minimum transmission rate & 1 bps/Hz\\ \hline
	\end{tabular}
    \vspace{-0.4cm}                   
\end{table}

\begin{figure}[t]
	\centering
	\includegraphics[width=2.5in]{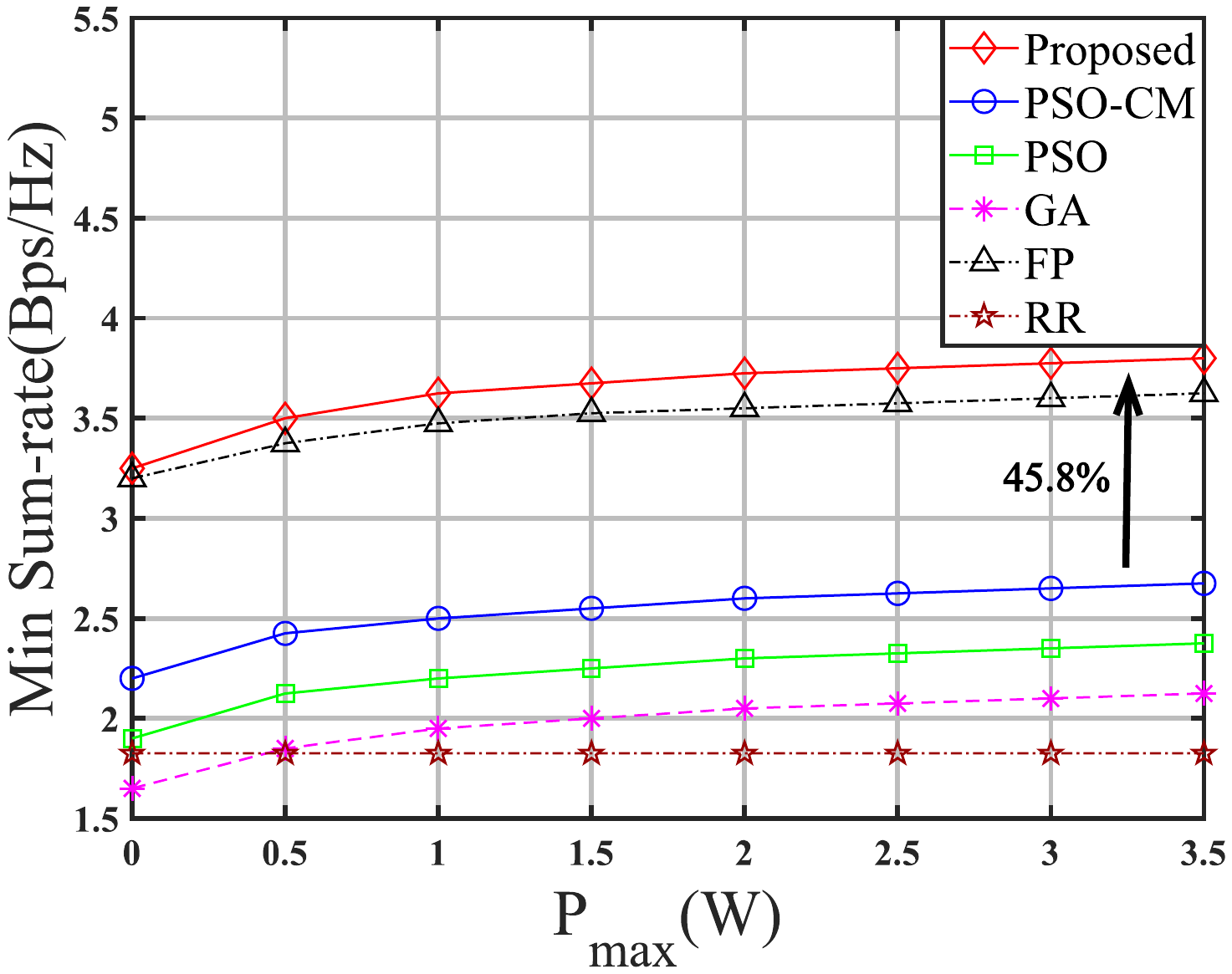}
	\caption{minimum sum-rate v.s. varying transmit power.} \label{F3b}
\end{figure}

Fig.~\ref*{F3b} examines the minimum sum-rate performance under varying maximal power constraints \(P_{\max}\). It shows that our proposed algorithm outperforms baseline algorithm PSO-CM by as large as $45.8\%$ in minimum sum-rate using $100$ iterations when $P_{\max} = 3.5$W. 
Compared with FP, Fig.~\ref*{F3b}  shows that sum-rate performance improvement becomes marginal when $P_{\max}>1\,\text{W}$. On the contrary, for \(P_{\max}=[0.5, 1, 1.5, 2, 2.5, 3]\text{W}\), our average power consumption is \([0.39, 0.69, 0.99, 1.38, 1.70, 2.01]\text{W}\), saving \([22\%, 31\%, 34\%, 31\%, 32\%, 33\%]\) transmit power. This suggests that maximum power is unnecessary for the required rate, as higher power increases interference as well. 
Overall, our transmit power optimization is shown to be effective.

\begin{figure}[t]
	\centering
	\includegraphics[width=2.3in]{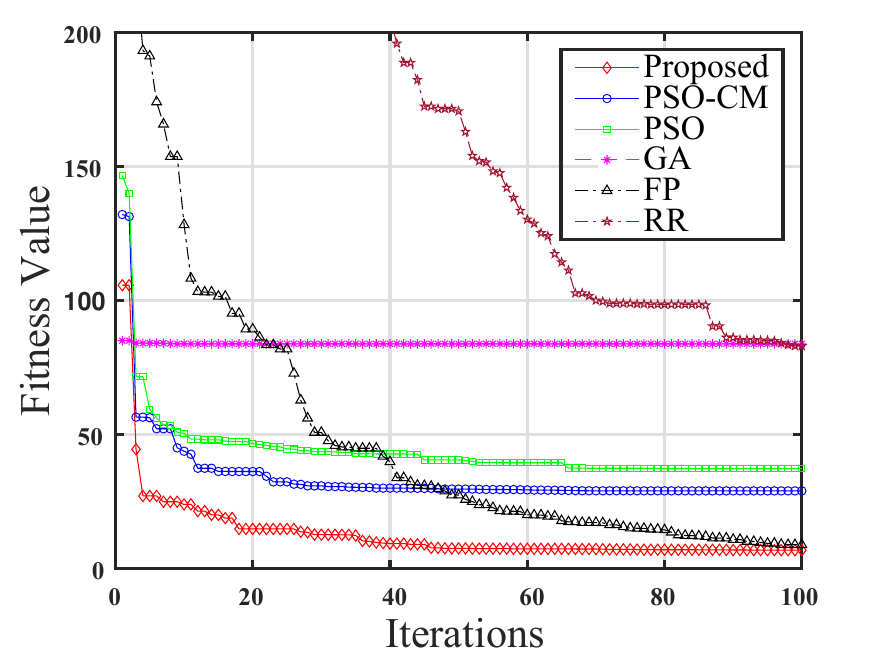}
	\caption{Fitness value v.s. number of iterations.} \label{F3c}
\end{figure}

Fig.~\ref*{F3c} validates that both the proposed and baseline algorithms converge, with our SCA-based particle initialization warm-up strategy enabling faster convergence and significantly better performance than the baseline. This shows the power of WS-PSO-CM over baselines. Furthermore, by numerical validation, solutions to Problem $(\textsc{P}1)$ and Problem  $(\textsc{P}2)$ are not KKT points in our settings.

\begin{figure}[t]
	\centering
	\includegraphics[width=2.5in]{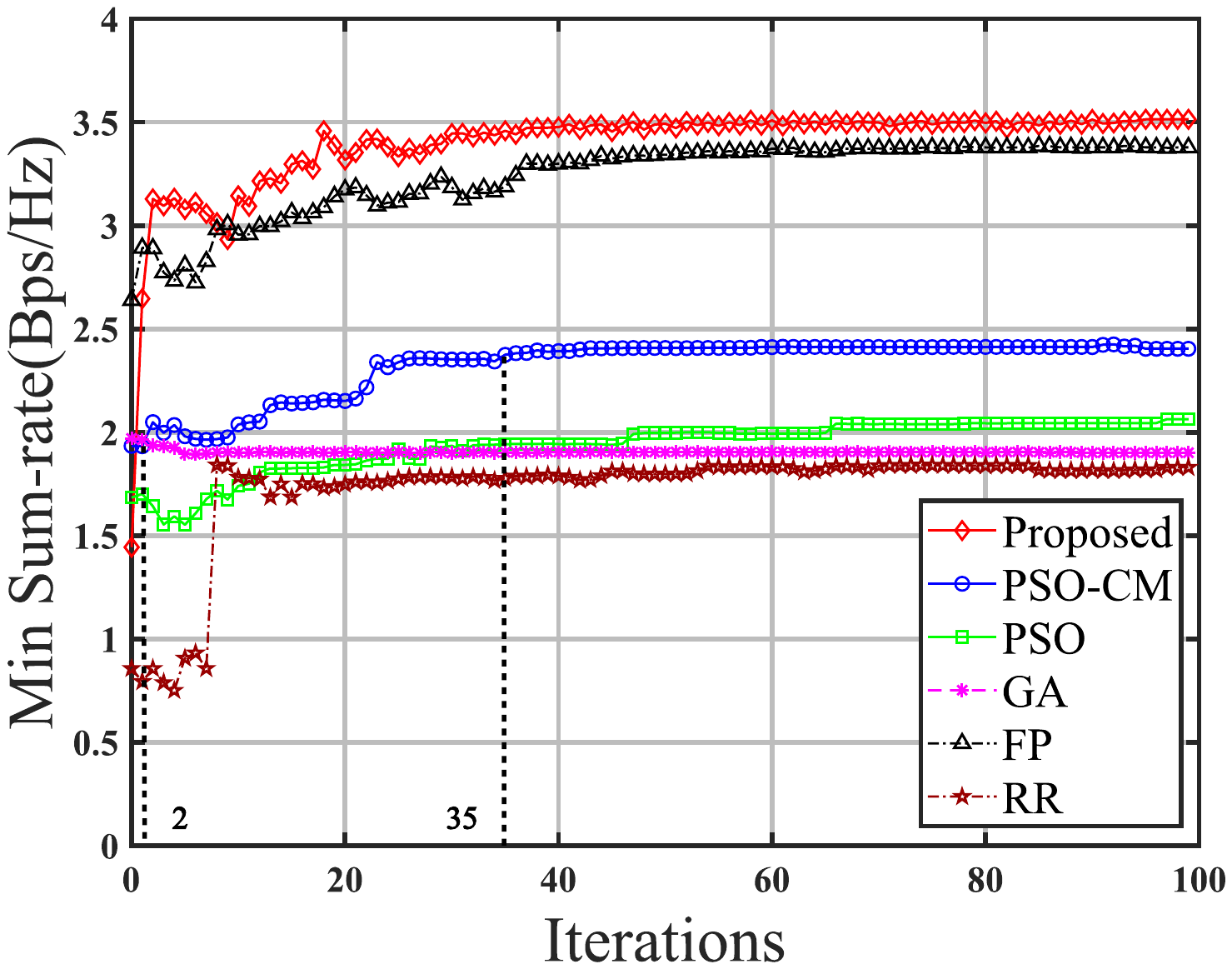}
	\caption{minimum sum-rate v.s. number of iterations.} \label{F3d}
\end{figure}

In Fig.~\ref*{F3d}, the proposed algorithm significantly improves the minimum sum-rate with $P_{\max} = 0.5$W. In simulation, the SCA warm-start only requires around $50$ seconds, while PSO-CM or WS-PSO-CM requires $5$-$6$ minutes. This implies the algorithm achieves a much higher minimum sum-rate with only marginal computation time added. In particular, Fig.~\ref*{F3d} shows that if WS-PSO-CM achieves the same minimum sum-rate with PSO-CM, only $2$ iterations are needed rather than $35$ iterations. Equivalently, this shows that WS-PSO-CM reduces $46.7\%$ computational time of PSO-CM. This is because the model-based particle initialization guides particles toward a favorable solution space. In addition, RR algorithm suffers from the worst minimum sum-rate among all algorithms, since RR algorithm prevents UAVs from approaching the nearest UGVs for achieving a higher rate. 
		
		
		

	\section{Conclusion}	
	
		In this paper, we studied the joint optimization of 3D trajectory and  communication in low-altitude air-ground cooperation systems, which would be a key application of low-altitude economy. We proposed an algorithm combining SCA-based scheduling, power control, and a novel WS-PSO-CM method for trajectory optimization, where WS-PSO-CM largely outperforms PSO-CM in \cite{PanTCOM}. By leveraging the environmental-aware radio map, we reveal that UAVs tend to avoid interference by buildings and fly near UGVs to mitigate PL.
		
		


		\bibliographystyle{IEEEtran}
		\bibliography{GC2025Ref}
		
	
	\end{document}